\documentclass[conference]{IEEEtran}
\IEEEoverridecommandlockouts
\usepackage[T1]{fontenc}
\usepackage{stfloats}
\usepackage{amsmath}
\usepackage{amsfonts}
\usepackage{bbding}
\usepackage{amssymb}
\usepackage{array}
\usepackage{subfigure}
\usepackage{graphicx}
\usepackage{subfigure}
\usepackage[named]{algo}
\usepackage{psfrag}
\usepackage{xfrac}
\usepackage{booktabs}
\usepackage{cases}
\usepackage[compress]{cite}
\usepackage{hyperref}
\usepackage{bm}
\usepackage{multirow}
\usepackage{amsthm}
\usepackage{stfloats}
\usepackage{setspace}
\usepackage{color}
\usepackage{algorithm}
\usepackage{algorithmicx}
\usepackage{algpseudocode}
\usepackage{cleveref}
\usepackage{makecell}

\makeatletter
\renewcommand{\citepunct}{,\penalty\@m\hskip.13emplus.1emminus.1em}
\renewcommand{\citedash}{\hbox{--}\penalty\@m}

\makeatother

\allowdisplaybreaks

\begin{document}
\title{Federated Learning Based Proactive Handover in Millimeter-wave Vehicular Networks}

\author{
\IEEEauthorblockN{Kaiqiang Qi, Tingting Liu and Chenyang Yang}
\thanks{This work is supported by National Natural Science Foundation of China (NSFC) under Grant 61671036.}
\IEEEauthorblockA{Beihang University, Beijing, China\\Email: \{kaiqiangqi, ttliu, cyyang\}@buaa.edu.cn}}
\maketitle
\begin{abstract}
Proactive handover can avoid frequent handovers and reduce handover delay, which plays an important role in maintaining the quality of service (QoS) for mobile users in millimeter-wave vehicular networks. To reduce the communication cost of training the learning model for proactive handover, we propose a federated learning (FL) framework. The proposed FL framework can accommodate the limited storage capacity of each user, increase the number of users who participate in the FL, and adapt to the dynamic mobility pattern. Simulation results validate the effectiveness of the proposed FL framework. Compared to reactive handover schemes, the proposed handover scheme can reduce the unnecessary handovers and improve the QoS of users simultaneously.
\end{abstract}


\section{Introduction}\label{Section:Introduction}
Millimeter-wave (mmWave) vehicular communication can support the high-capacity demands of mobile users in the fifth generation (5G) cellular systems \cite{Choi2016Millimeter}. However, the special propagation characteristics at the mmWave band, say high susceptibility to blockages, can easily incur the connection intermittency \cite{Giordani2017Millimeter}. This makes mobility management challenging, especially the handover management for maintaining the quality of service (QoS) of vehicular users.

Prevalent handover schemes are reactive, which are initiated only if the triggering condition will be fulfilled for a duration of Time-to-Trigger (TTT) \cite{Lopez2012Mobility}. A short TTT causes unnecessary handovers and ping-pong effect, whereas a long TTT postpones the handover, leading to the increasing probability of handover failures. To improve the QoS during handover, proactive handover schemes were proposed, which harnessed the historically observed data by machine learning \cite{I2017bigdata,Wang2018Handover,Sun2018SMART,Alkhateeb2018Machine,Lee2020CHO}. In \cite{I2017bigdata}, a support vector machine based proactive handover method was proposed to decrease the numbers of service interruption and impact of ping-pong effect in Long-Term Evolution networks. In \cite{Wang2018Handover}, the handover decision was learned via deep reinforcement learning to avoid frequent handovers in ultra-dense networks. To circumvent the severe impacts of the sudden signal attenuation caused by blockages, proactive handover schemes were investigated for mmWave networks considering various learning techniques, say multi-arm bandit, gate recurrent unit, and deep neural networks \cite{Sun2018SMART,Alkhateeb2018Machine,Lee2020CHO}.

All these works resort to centralized learning \cite{Sun2018SMART,Alkhateeb2018Machine,Lee2020CHO}, where the users need to upload their data to a base station (BS) or a central processor (CP) for training, which may incur high communication cost. Besides, if user locations are required for training, some users may be unwilling to share their data with the BS or CP due to privacy consideration. A distributed training framework, federated learning (FL), can exploit the local data and computation resources at the users and make the mobile equipment more intelligent. While FL has attracted considerable attention in wireless community, most research efforts are devoted to improve the learning performance in wireless systems whereas much fewer works study how to improve communication performance with FL \cite{Lim2020Federated}. To address the privacy issue, FL was investigated to predict the content popularity for proactive caching at the wireless edge in \cite{Qi2020Popularity}. To reduce the communication cost, FL was developed for the hybrid beamforming in mmWave systems in \cite{Elbir2020Beamforming}, and for distributed power allocation to maximize the energy/spectrum efficiency in \cite{Yan2020Federated}. Most existing works either consider non-mobile users or a fixed group of users participating in the FL.

In vehicular networks, however, users arrive at the coverage of a BS or CP asynchronously and with random sojourn time. Besides, each user cannot or is not willing to store a large amount of historical data due to the limited storage capacity. As a consequence, using FL for proactive handover encounters the following three major challenges:
\begin{enumerate}
  \item The available local training data is limited by the storage capacity of users.
  \item The users to participate in the FL arrive asynchronously.
  \item The FL should adapt to the dynamically changing mobility pattern.
\end{enumerate}

In this paper, we propose a FL framework for proactive handover in the mmWave vehicular networks. We consider heterogeneous networks, where each macro BS (MBS) serves as a CP.  To address the three challenges, we introduce streaming, asynchronous and online training into the FL framework to predict the next associated small BS (SBS). In particular:
\begin{enumerate}
\item The \emph{streaming FL} can leverage the limited storage space at each user during local model updating.
\item The \emph{asynchronous FL} can allow more users to participate in the distributed training so as to improve performance by global model aggregation.
\item The \emph{online training with transfer learning} can refine the model with high efficiency when user mobility pattern  changes dynamically.
\end{enumerate}

Simulation results validate the effectiveness of the proposed FL framework.

The remainder of the paper is organized as follows. Section II describes the system model and the proactive handover scheme. Sections III introduces the FL framework for proactive handover. Section IV provides simulation results. Finally, we conclude this paper in Section V.

\section{System Model and Proactive Handover}
Consider a mmWave vehicular network with multiple regions, where each region is covered by a MBS operated in low frequency band and $N_{\mathrm b}$ mmWave SBSs. Let $\mathcal{B}=\{1,\cdots,N_{\mathrm b}\}$ be the set of the SBSs in a region.

Let a frame with duration $T_{\mathrm{f}}$ (say 100 ms) denote a handover measurement period. The average channel gain is assumed staying constant in each frame but may vary among different
frames.
In the $f$th frame, the average signal-to-noise ratio (SNR) of the $k$th user received from the $b$th SBS is
\begin{align}
  \gamma_{b,k}^f=\frac{P_\mathrm{tx}G\alpha_{b,k}^f}{\sigma_{\mathrm N}^2},
\end{align}
where $P_{\mathrm{tx}}$ is the transmit power of the SBS, $G$ is the beamforming gain of the SBS,  $\alpha_{b,k}^f$ is the average channel gain of the $k$th user from the $b$th SBS in the $f$th frame, and $\sigma_{\mathrm N}^2$ the noise power.




At the end of each frame, say the $f$th frame, the user judges whether the handover trigger condition is satisfied, i.e.,
\begin{equation}\label{eq:TrigCondition}
\gamma_{b_k^f,k}^f<\gamma_{\mathrm{th}}, \quad b_k^f\in \mathcal{B},
\end{equation}
where $b_k^f$ denotes the index of the associated SBS of the $k$th user in the $f$th frame, and $\gamma_{\mathrm{th}}$ is the expected SNR of the user, which can reflect the user's QoS requirement.

Once the trigger condition in \eqref{eq:TrigCondition} is satisfied, each user initiates a handover. A reactive handover scheme with TTT of duration $T_\mathrm{p}$ can decrease the number of unnecessary handovers and avoid the ping-pong effect, but suffers from the handover delay with $N_\mathrm{p} = T_\mathrm{p}/T_\mathrm{f}$ frames.
On the other hand, a reactive scheme without TTT can avoid the handover delay, but suffers from the ping-pong effect.

In the sequel, we investigate a proactive handover scheme to reduce the number of unnecessary handovers and the handover delay simultaneously.
In particular, each user chooses the next associated SBS, denoted by $\hat{b}^{\mathrm{next}}_k$, according to the predicted SNRs in a prediction window, where the prediction is based on the measured SNRs in an observation window. The prediction window contains the subsequent $N_{\mathrm p}$ frames after the user initiates a handover, while the observation window contains the preceding $N_{\mathrm o}$ frames. The SNR vectors in the prediction window and observation window are $\hat{\bm{\gamma}}^{\mathrm{pre}}_{b,k}=[\hat{\gamma}^{f+1}_{b,k},\cdots,\hat{\gamma}^{f+N_{\mathrm p}}_{b,k}]$ and $\bm{\gamma}^{\mathrm{obs}}_{b,k}=[\gamma^{f-N_{\mathrm o}+1}_{b,k},\cdots,\gamma^f_{b,k}]$ respectively, where $\hat{{\gamma}}_{b,k}^f$ is the predicted value of ${\gamma}_{b,k}^f$.

To improve the QoS during the handover, the $k$th user can select the next associated SBS providing the maximum average predicted SNR during the prediction window, i.e., $\bar{{\gamma}}^{\mathrm{pre}}_{b,k}=\sum_{i=1}^{N_{\mathrm p}}\hat{\gamma}_{b,k}^{f+i}/{N_{\mathrm p}}$. However, if the values of ${\bm{\gamma}}^{\mathrm{pre}}_{b,k}$ for the SBS with the largest $\bar{{\gamma}}^{\mathrm{pre}}_{b,k}$ fluctuate significantly, some SNRs in the prediction window are probably less than $\gamma_{\mathrm{th}}$, leading to frequent handover triggers. To avoid unnecessary triggers, we first select some SBSs whose SNRs in the prediction window are always no less than $\gamma_{\mathrm{th}}$, i.e.,
\begin{equation}\label{eq:Candidate}
\mathcal{B}_k^f = \{b|\hat{\gamma}_b^{f_0}\ge\gamma_{\mathrm{th}},\forall f_0=f+1,\cdots,f+N_{\mathrm p}\}.
\end{equation}

If $\mathcal{B}_k^f$ is nonempty, the next associated SBS $\hat{b}^{\mathrm{next}}_k$ is selected from $\mathcal{B}_k^f$. Otherwise, $\hat{b}^{\mathrm{next}}_k$ is selected from the all SBSs. To avoid unnecessary handovers, the user keeps the associated SBS unchanged during the prediction window, i.e.,
\begin{equation}
b_k^{f_0} = \hat{b}_k^{\mathrm{next}}, ~f_0=f+1,\cdots,f+N_{\mathrm p},
\end{equation}
where
\begin{equation}
\hat{b}^{\mathrm{next}}_k =
\begin{cases}
\mathop{\arg\max}_{b\in{\mathcal{B}_k^f}}\bar{{\gamma}}^{\mathrm{pre}}_{b,k},\quad&\text{if }\mathcal{B}_k^f\neq\emptyset,\\
\mathop{\arg\max}_{b\in{\mathcal{B}}}\bar{{\gamma}}^{\mathrm{pre}}_{b,k},\quad&\text{otherwise}.
\end{cases}
\end{equation}
If the next associated SBS is different from the currently associated SBS, i.e., $\hat{b}^{\mathrm{next}}_k \neq b_k^f$, the user starts the handover. Otherwise, the user is still associated with the current SBS in the following ${N_{\mathrm p}}$ frames.

\section{Federated Learning for Proactive Handover}
In this section, we introduce the training process of FL for proactive handover, consisting of multiple communication rounds. In each round, each passing user who participates in FL first uses the previously stored training samples to update a local model, and then the MBS aggregates the uploaded models as a global model.

We consider supervised learning to implement the proactive handover. To reduce the number of unnecessary prediction parameters, we predict the next associated SBS, $\hat{b}^{\mathrm{next}}_k$, instead of the SNR vectors in the prediction window. Then, for the $k$th user, the input vector of a training sample is $\bm{x}_k=[\bm{\gamma}^{\mathrm{obs}}_{1,k},\cdots,\bm{\gamma}^{\mathrm{obs}}_{N_{\mathrm{b}},k}]$, which is a $N_{\mathrm b} N_{\mathrm o}$-dimension vector, and the expected output vector of the training sample (i.e., the label), $\bm{y}_k$, is a $N_{\mathrm b}$-dimension one-hot vector, where the index of the nonzero element indicates the next associated SBS.

\subsection{Local Model Update at Each User}
In the FL framework, the model is trained with  $T$  communication rounds between the MBS and the users. In each communication round, say the $t$th round, the MBS broadcasts the  global model  (denoted as $\mathbf{w}_{\mathrm g}^t$) multiple times, to ensure that every passing user receives the model and hence has the opportunity to participate in the FL. Then, each participant user downloads the global model from the MBS, and initializes the local model with the global model, i.e., $\mathbf{w}_k^t = \mathbf{w}_{\mathrm g}^t$.

The $k$th participant user trains its local model to minimize the loss function, i.e.,
\begin{align}\label{eq:LocalModel}
  \mathbf{w}_k^{t*}=\mathop{\arg\min}\left\{L(\mathbf{w}_k^t)\right\},
\end{align}
where $L(\mathbf{w}_k^t)$ is the cross-entropy (since the problem at hand is a classification problem) between the output vector of the model $\hat{\bm{y}}_k^{(i)}$ and the expected output vector $\bm{y}_k^{(i)}$
\begin{align}\label{eq:cross-entropy}
L(\mathbf{w}_k^t)=-\frac{1}{|\mathcal{D}_k|}\sum\nolimits_{i\in\mathcal{D}_k}\sum\nolimits_{b=1}^{N_{\mathrm b}}y_{k,b}^{(i)}\log{\hat{y}_{k,b}^{(i)}},
\end{align}
where $|\!\cdot\!|$ is the cardinality of a set, $\mathcal{D}_k =\{(\bm{x}_k^{(i)},\bm{y}_k^{(i)})\}$ is the local training set of the $k$th participant user, and $\bm{x}_k^{(i)}$ and $\bm{y}_k^{(i)}$ are respectively the input vector and the expected output vector of the $i$th training sample.

In \eqref{eq:LocalModel}, the model parameters can be updated by back-propagation with the stochastic gradient descent (SGD) algorithm. To reduce the communication cost, the local update can be performed several epoches through the training set $\mathcal{D}_k$ as the Federated Averaging algorithm \cite{Mcmahan2017communication}.

The performance of local models  depends on the available training data in $\mathcal{D}_k$, which is limited by the storage capacity of the users. In practice, a user may traverse a region frequently, which can obtain a large number of local training samples for proactive handover. To improve the performance of local model,  \emph{traditional FL} assumes that each user has sufficiently large storage space to store all the historical samples. However, the user may traverse many regions covered by different MBSs each day. It is not affordable for mobile equipments to store the gathered data in all regions of many days, which may affect the willingness of users for participating in the FL severely.

Considering the limited storage space of mobile users, we propose \emph{streaming FL}, where each user stores the training samples in a streaming style. Specifically, when a user enters a region, the local training set only contains a part of samples generated in this region along the user's movement route. The training samples will be discarded, after the samples have been used for training or when the user enters into another region.

\subsection{Global Model Aggregation at MBS}

After all participant users upload their updated local models $\mathbf{w}_k^{t}$ to the MBS, the MBS aggregates the local models as the global model in the $(t+1)$th round, i.e.,
\begin{equation}\label{eq:Aggregation}
	\begin{aligned}
	\mathbf{w}_{\mathrm g}^{t+1} = \sum\nolimits_{k\in \mathcal{K}^t}\omega_k\mathbf{w}_k^t,
	\end{aligned}
\end{equation}
where $\omega_k = |\mathcal{D}_k|/\sum_{m\in\mathcal{K}^t}|\mathcal{D}_m|\in[0,1]$ is the aggregation coefficient, i.e., the fraction of training samples of the $k$th user, and $\mathcal{K}_t$ is the participant user set in the $t$th round.

In the FL framework, there are two ways for users to upload their local models, i.e., \emph{synchronous FL} and \emph{asynchronous FL}. With synchronous FL, the communication cost can be reduced significantly by using over-the-air computation \cite{Yang2020Federated}, where all users transmit their local models to the MBS simultaneously.
The synchronous FL requires the participant users to meet the following conditions: 1) appear in one region simultaneously; 2) are willing to participate in FL, and have training data and enough energy to complete the local computation; 3) maintain very accurate time synchronization in millisecond or even microsecond-level to upload local models. However, due to the random arrival of mobile users, the number of users who can satisfy the above conditions is very limited. Since the performance of FL improves with the number of participant users $|\mathcal{K}_t|$ \cite{Yang2020Federated}, the performance of synchronous FL may be unsatisfactory due to few participant users.

To allow more users participating in FL in each round, we consider \emph{asynchronous FL}. With the asynchronous FL, each user uploads the model immediately after the local update. Once receiving the updated model from the user, the MBS iteratively aggregates the current model of this round in an asynchronous manner. By increasing the duration of each communication round $T_\mathrm{r}$, the MBS can wait for more users to arrive such that the number of participant users increases significantly.

Specifically, when the MBS has received the local model from the $k$th participant users, it updates the global model as follows,
\begin{equation}\label{eq:AsyAggreagtion}
	\begin{aligned}
	\mathbf{w}_{\mathrm{g},k}^t &=
		\begin{cases}
		\mathbf{w}_{k}^t,&\text{if }k=1,\\
		\beta_k\mathbf{w}_{k}^t+(1-\beta_k)\mathbf{w}_{\mathrm{g},k-1}^t, &\text{if }1<k\le|\mathcal{K}^t|,
		\end{cases}
	\end{aligned}
\end{equation}
where $\beta_k=|\mathcal{D}_k|/\sum_{m=1}^{k}|\mathcal{D}_m|\in[0,1]$ is a weighted coefficient, which is proportional to the number of training samples of the $k$th user among all the participant users.

If the deadline for each round is due, or a user has left this region, then the updated model of the user is unnecessary to be uploaded. When the $t$th communication round has terminated, from \eqref{eq:AsyAggreagtion}, we can obtain the global model finally aggregated at the MBS as $ \mathbf{w}_{\mathrm g}^{t+1} = \mathbf{w}_{\mathrm{g},|\mathcal{K}^t|}^t =\sum\nolimits_{k\in \mathcal{K}^t}\omega_k\mathbf{w}_k^t$, which is the same as \eqref{eq:Aggregation}.

\subsection{Online Training With Transfer Learning}
The training process is usually performed in an offline manner, where each participant user only contributes to the local model update, but does not use the aggregated model for handover prediction until the global model has converged. The offline-trained model can work well if user mobility pattern remains unchanged.

In vehicular networks, however, the user mobility pattern changes inevitably. For example, the average velocities of users in a region change from peak time to off-peak time. Then, the model has to be re-trained whenever it cannot be generalized to the new scenario. In order to be adaptive to the new mobility pattern, we resort to \emph{online training} by refining the model. To improve the learning efficiency  and speed up the convergence of FL, we consider \emph{transfer learning}, i.e., use the offline-trained model as the initialization of the online-trained model.

To implement the online training, a part of users are scheduled to participate in the FL. In the first round of the online training phase, the global model at the MBS is initialized as $\tilde{\mathbf{w}}_{\mathrm g}^1 = \mathbf{w}_{\mathrm g}^{T+1}$, where $\mathbf{w}_{\mathrm g}^{T+1}$ is the offline-trained model. In the $t$th round of the online training phase, when the handover is triggered according to \eqref{eq:TrigCondition}, by inputting the SNR vector in the observation window, the user predicts the next associated SBS using the online-trained model (i.e., $\tilde{\mathbf{w}}_{\mathrm g}^{t}$), which is broadcasted by the MBS. The training procedure for the online training is the same as that for the offline training, but the durations of a communication round may be different in the two phases.



\section{Simulation Results}
In this section, we evaluate the performance of the proposed FL based proactive handover scheme.


\subsection{Simulation Setups and Data set}
\begin{figure}[!htb]
	\centering
    \begin{minipage}[t]{0.47\textwidth}
    \includegraphics[width=1\textwidth]{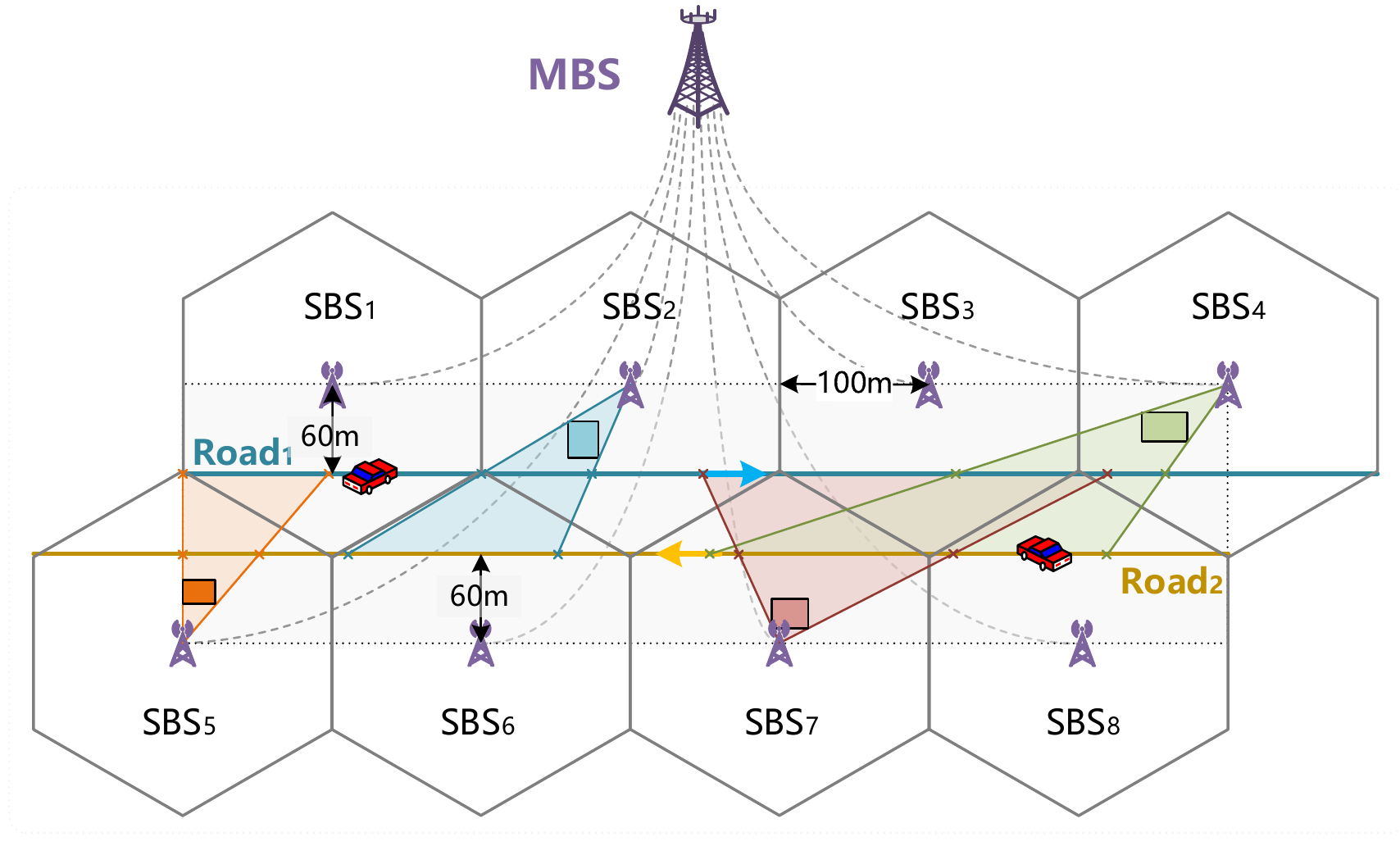}
    \end{minipage}
	\caption{Considered road and network topology in the mmWave vehicular system, where four rectangular obstacles (say the buildings) are randomly distributed at both sides of the two roads.}
	\label{fig:Road_Topology}
\end{figure}

Consider a heterogeneous network consisting of a MBS and $N_{\mathrm b}=8$ mmWave SBSs, where each SBS with radius of $100$~m is located in the center of a small hexagonal cell as shown in Fig. \ref{fig:Road_Topology}. The vehicles travel along two straight roads with right or left direction, and both roads are with minimum distance of $60$ m from the SBSs.
The initial velocities of the users are uniformly distributed between $15$ m/s and $20$ m/s. The acceleration of each user is Gaussian random variable with standard deviation $1\text{ $\mathrm{m/s^2}$}$.
Other simulation parameters are listed in Tab. \ref{tab:Simu_Para}~\cite{Akdeniz2014Millimeter,3GPP38901,Sun2018SMART}.

\begin{table}[htb!]
 \centering
  \vspace{-1mm}
 \caption{Simulation Parameters}\label{tab:Simu_Para}
\footnotesize
 \begin{tabular}{l|l}
  \hline
  \textbf{Parameters} & \textbf{Values} \\
  \hline
  \hline
  Pathloss (LOS) & $61.4+20\log_{10}(d)$ dB    \\
  \hline
  Pathloss (NLOS) & $72.0+29.2\log_{10}(d)$ dB    \\
  \hline
  Standard derivation of shadowing (LOS) &  $5.8$ dB    \\
  \hline
  Standard derivation of shadowing (NLOS) &  $8.7$ dB    \\
  \hline
  Correlation distance of shadowing (LOS)  & $10$ m  \\
  \hline
  Correlation distance of shadowing (NLOS)  & $13$ m  \\
  \hline
  Duration of a frame, $T_{\mathrm f}$ & $100$ ms \\
  \hline
  Transmit power of each SBS, $P_{\mathrm{tx}}$ & 30 dBm\\
  \hline
  Beamforming gain of each SBS, $G$ & 18 dB\\
  \hline
  Noise power, $\sigma_{\mathrm N}^2$ & -77 dBm\\
  \hline
  Expected SNR of each user, $\gamma_{\mathrm{th}}$ & $22$ dB\\
  \hline
  Length of observation window, $N_{\mathrm o}$ & $3$ frames\\
  \hline
  Length of prediction window, $N_{\mathrm p}$ & $5$ frames\\
  \hline
  TTT, $T_{\mathrm p}$ & 500 ms\\
  \hline
\end{tabular}
 \vspace{-0.5mm}
\end{table}

To generate training and test samples for learning in the proactive handover, we first synthesize the trajectory sampled with duration of one frame. Then, we calculate the average channel gains according to the pathloss and shadowing models, where both the light-of-sight (LOS) and non-line-of-sight (NLOS) cases are considered. Finally, we obtain the SNR from each BS to each user.
To mitigate the impact of shadowing, the SNR in each frame is further processed by a moving averaging filter with coefficient of $0.5$ \cite{Lopez2012Mobility}. From each trajectory, we can obtain about $24$ samples. Each sample is $(\bm{x}_k^{(i)},\bm{y}_k^{(i)})$, where $\bm{x}_k^{(i)}$ is the observed SNR vector and $\bm{y}_k^{(i)}$ is the one-hot vector to denote the next associated SBS.

The samples are divided into two groups, where the samples in the 1st group and 2nd group are generated by the users moving along Road$_1$ and Road$_2$, respectively. As shown in Fig. \ref{fig:Road_Topology}, the users whose samples are in the 1st group may associate with the SBSs in the first row, while the others whose samples are in the 2nd group are likely to associate with the SBSs in the second row. As a result, the associated SBSs for the two groups of samples are different. Moreover, the movement direction and channel environments in different groups are also different. Consequently, the samples in different groups are non-independent identically distributed (Non-IID).
The test set contains $24,000$ samples, where one half of samples are from the 1st group and the other are from the 2nd group. If the training samples are from one group, the test samples and training samples are Non-IID, otherwise, they are IID.

For the FL, we need to analyze the number of participant users in each round of the region. In an urban region, the average vehicle arrival rate during daytime is $a=50$ vehicles/minute/road \cite{Tian2018LSTM}. Considering that the length of each road is $800$ m and the average velocity is $20$ m/s, the average sojourn time of each user in the road is $T_{\mathrm{soj}}=40$ s. For the synchronous FL, the number of users who move in the two roads simultaneously is no more than $2aT_{\mathrm{soj}}\approx67$. Since not all the users are willing to participate in the FL, the number of participant users is very limited. By contrast, for the asynchronous FL, given the duration of each communication round $T_\mathrm{r}$, the number of passing users in each round is $2aT_{\mathrm r}$. We can increase the number of participant users by setting $T_\mathrm{r}$ as a large value. For example, if $T_{\mathrm r}=1$ minute, there are $2aT_{\mathrm r}=100$ users. In the following, we consider the performance of the asynchronous FL with $T_{\mathrm r}=1$ minute.

\subsection{Performance of Federated Learning}
We adopt the multi-layer perception as an illustrative machine learning model. The fine-tuned hyper-parameters of the multi-layer perceptron are listed in Tab. \ref{tab:NN_Para}, which are used for both centralized and federated learning.
\begin{table}[htb!]
 \centering
  \vspace{-1mm}
  \caption{Fine-tuned Hyper-Parameters of the Multi-layer Perceptron}\label{tab:NN_Para}
  \footnotesize
 \begin{tabular}{l|l}
  \hline
  \textbf{Hyper-parameters} & \textbf{Values} \\
  \hline
  \hline
  Number of input nodes & $24$ \\
  \hline
  Number of hidden layers & $1$    \\
  \hline
  Number of hidden nodes & $20$ \\
  \hline
  Number of output nodes & $8$ \\
  \hline
  Activation function of hidden layers & Sigmoid\\
  \hline
  Activation function of output layer & Softmax\\
  \hline
  Learning algorithm & SGD \\
  \hline
 \end{tabular}
 \vspace{-0.5mm}
\end{table}
\subsubsection{Accuracy of Federated Learning}
To provide an upper bound of performance for FL, we provide the accuracy of centralized learning. For the centralized learning, $1,400$ training samples are required to achieve the steady test accuracy of $95.58\%$, where the test accuracy is computed as the ratio of samples with right prediction in the test set, the hyper-parameters are fine-tuned with learning rate of $0.05$ and number of epoches of $2,000$.
\begin{figure*}[!htb]
	\centering
    \subfigure[]{
        \begin{minipage}[b]{0.31\textwidth}
        \includegraphics[width=1\textwidth]{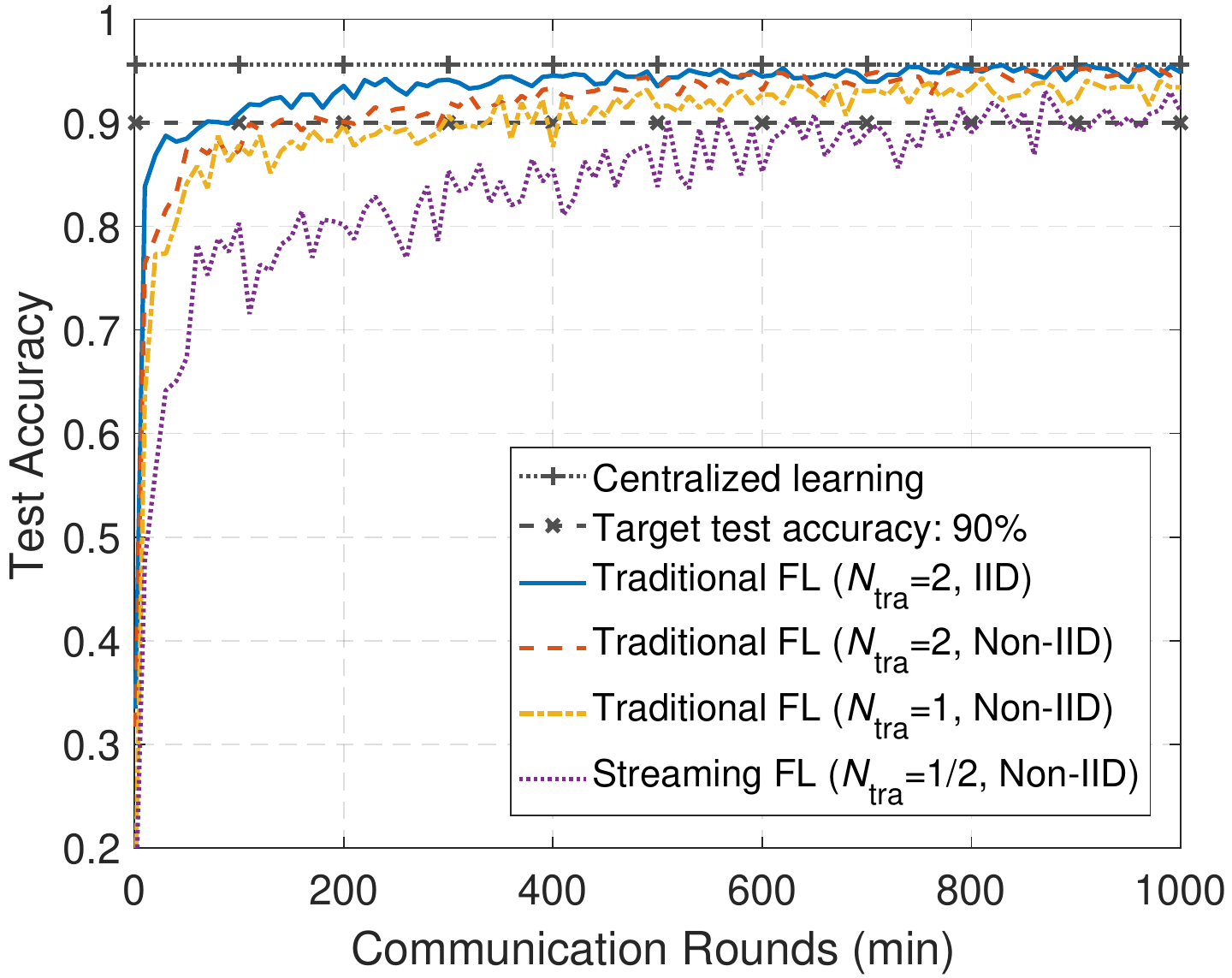}
        \end{minipage}
        }
    \subfigure[]{
        \begin{minipage}[b]{0.31\textwidth}
        \includegraphics[width=1\textwidth]{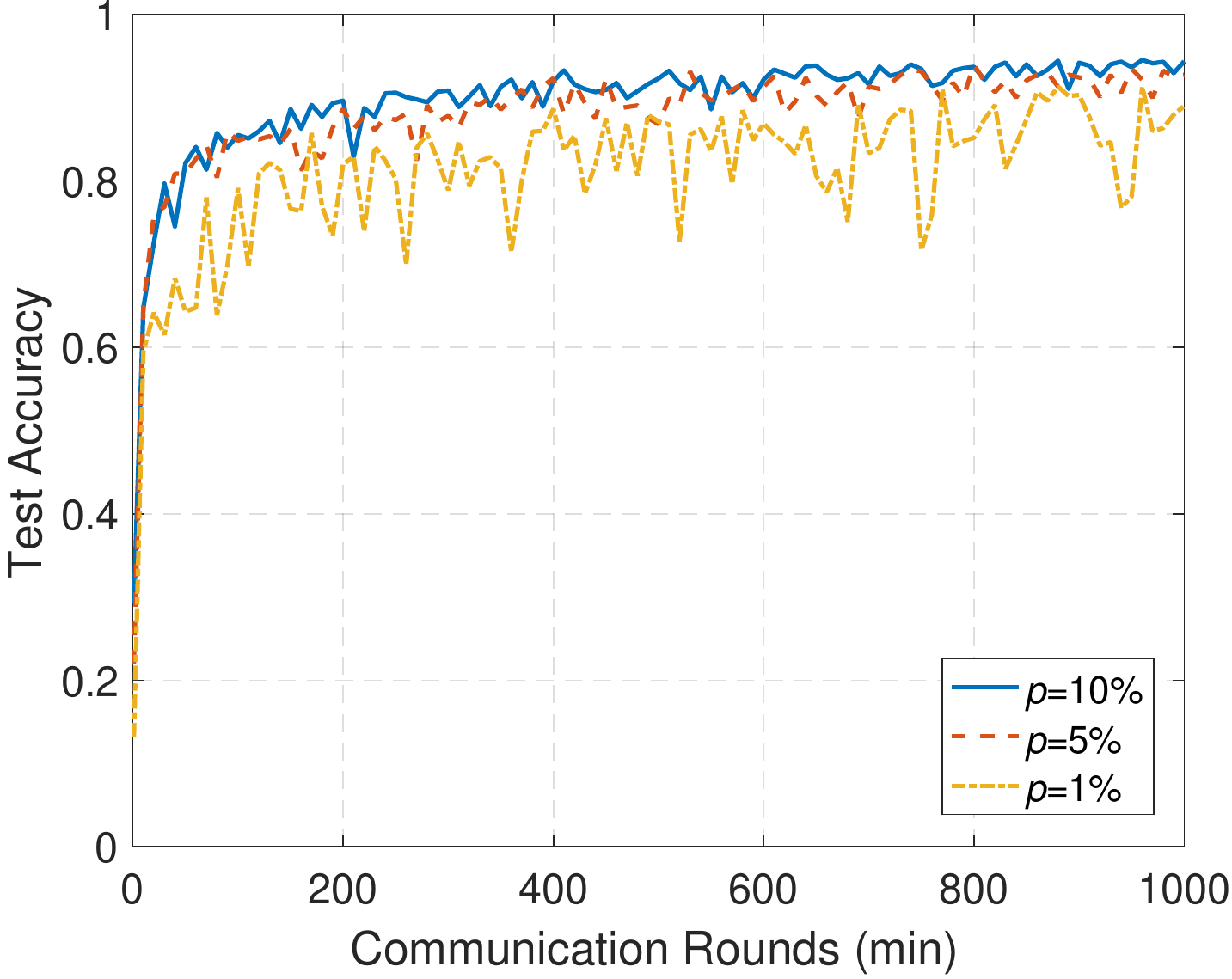}
        \end{minipage}
        }
    \subfigure[]{
        \begin{minipage}[b]{0.31\textwidth}
        \includegraphics[width=1\textwidth]{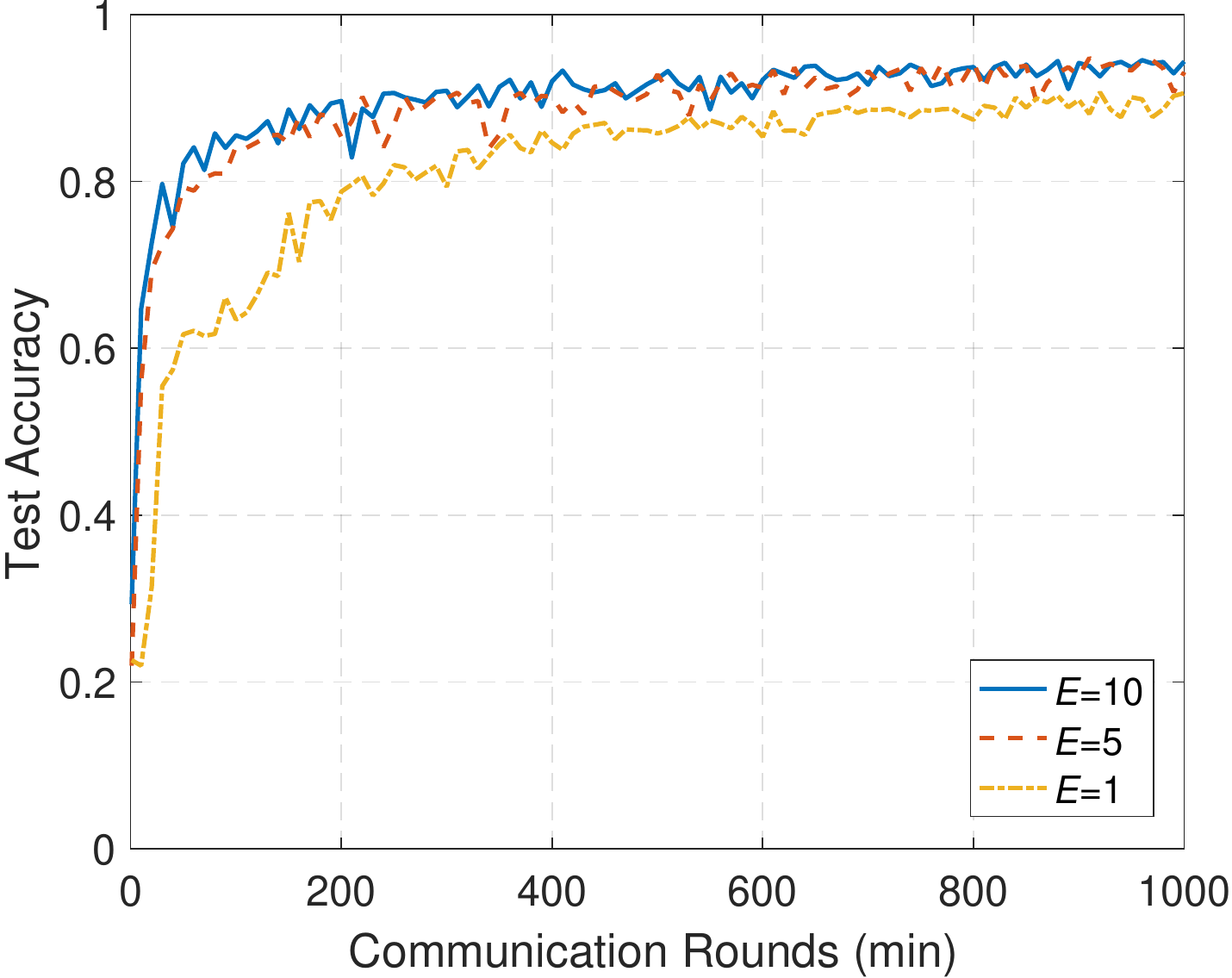}
        \end{minipage}
        }
	\caption{Impacts on the performance of the FL (a) Amount of stored data at each user, $p = 10\%$, and $E = 10$. (b) Fraction of participant users, $E = 10$, traditional FL with $N_{\mathrm{tra}}=1$. (c) Number of local update epoches, $p = 10\%$, $N_{\mathrm{tra}}=1$.}
	\label{fig:Convergence}
    \vspace{-0.5cm}
\end{figure*}

To show the impact of the amount of stored data at each user on the performance of the FL, we also compare the  streaming FL with the traditional FL in Fig. \ref{fig:Convergence}(a), where the fraction of users who are willing to participate in the FL in each round is $p=10\%$ (i.e., the average number of participant user is 10), and the number of epoches for local update in each round is $E=10$. We let $N_{\mathrm{tra}}$ denote the number of historical trajectories, from which the training samples stored at each user are obtained. For the streaming FL, since each user can only store the samples from a part of the trajectory, we set $N_{\mathrm{tra}}=1/2$ as an example. For the traditional FL, we set $N_{\mathrm{tra}}=1$ and $2$, respectively, and consider both the IID and non-IID settings.

As shown in Fig. \ref{fig:Convergence}(a), when each user stores the samples from two trajectories, i.e., $N_{\mathrm{tra}}=2$,  the test accuracy of the traditional FL can converge to the  steady test accuracy of the centralized learning. We can also see that for the traditional FL, storing more data can speed up the convergence, and the  convergence speed in the IID setting is slightly faster than the non-IID setting. The streaming FL converges slower than the traditional FL. To achieve a target test accuracy of $90\%$, the required minimum number of communication rounds for the streaming FL is about $800$, while that for the traditional FL with $N_{\mathrm{tra}}=1$ is about $300$. Nevertheless, when the user passes through multiple regions, the required storage space of the traditional FL increases with the number of regions, while the required storage space of the streaming FL does not increase. For example, if a user traverses $100$ regions, the required storage space of the traditional FL with $N_{\mathrm{tra}}$ is about $200$ times over that of the streaming FL with $N_{\mathrm{tra}}=1/2$.

In Figs. \ref{fig:Convergence}(b) and \ref{fig:Convergence}(c), we show the impacts of the fraction of participant users and the number of epoches for local update at each user in each round for the traditional FL. It can be observed that with more users participating in the FL, the convergence is faster with better steady test accuracy. This indicates that the asynchronous FL can outperform the synchronous FL. With the increase of the number of  epoches in each local update, the convergence of FL becomes faster. These conclusions are also applicable to the streaming FL.

\begin{figure}[!htb]
	\centering
    \begin{minipage}[t]{0.44\textwidth}
    \includegraphics[width=1\textwidth]{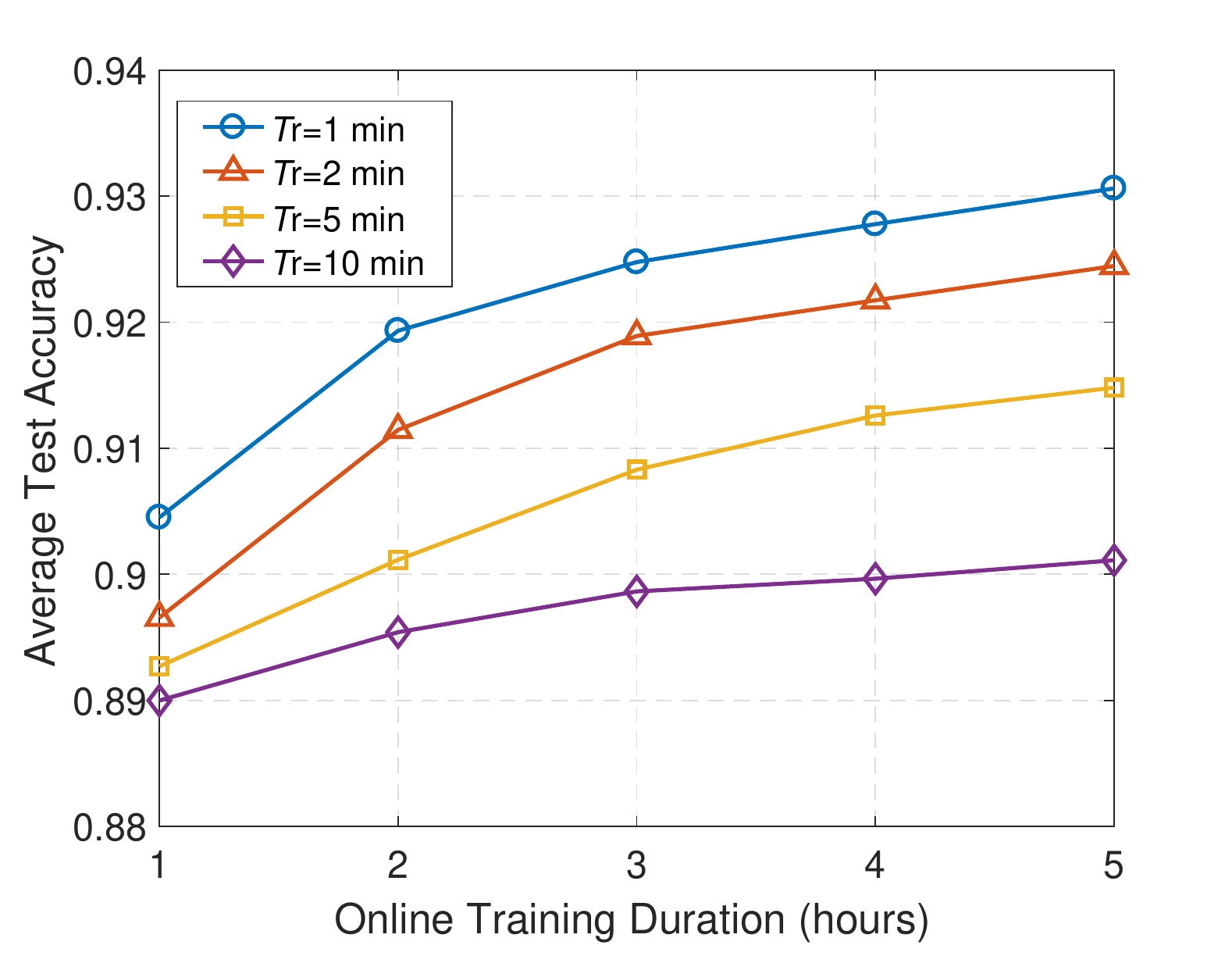}
    \end{minipage}
	\caption{Average test accuracy versus online training duration under different values of $T_{\mathrm r}$, $E=10$, and number of participant users in each round is $10$.}
	\label{fig:Online_Training}
  \vspace{-0.2cm}
\end{figure}
When the mobility pattern changes, for example, each user's initial velocity changes from $15\sim20$ m/s to $20\sim25$ m/s, it is necessary to employ the online training. In Fig. \ref{fig:Online_Training}, we show the performance of online training with transfer learning under different durations of a communication round (i.e., $T_\mathrm{r}$), where the streaming FL is considered in this phase. The performance metric is the average test accuracy during one hour. We can see that the average test accuracy can achieve more than $89\%$ in the first hour of online training, which indicates that the transfer learning is effective to speed up the convergence of FL. The performance can be further improved by increasing the number of communication rounds, e.g. by shortening $T_{\mathrm r}$ or lengthening the online training duration.

\subsubsection{Communication Costs}
We next compare the uplink communication costs of the centralized learning and FL based schemes, where the $16$-bit quantization for training data and model parameters are considered.

For the centralized learning, the measured SNRs of a user should be uploaded to the MBS when the user stays in the covered region. Hence, the amount of measured SNRs is about $N_{\mathrm b}\times T_{\mathrm{soj}}/T_{\mathrm f}\times16=51.2$ Kbits for $T_{\mathrm{soj}}=40$ s, and the average data rate required for uploading the measured data at each user is $51.2\text{K}/T_{\mathrm{soj}}=1.28$ Kbps. For the traditional FL, each participant user uploads the model parameters during $T_{\mathrm{soj}}$. The fine-tuned multi-layer perceptron contains $668$ parameters, including weights and biases. Then, the average data rate required for uploading the model parameters at each user is $(668\times16)/T_{\mathrm{soj}}\approx0.27$ Kbps. We can see that the FL is with less required bandwidth resources in the offline training phase.

\subsection{Performance of Handover}
We compare the proposed FL based proactive handover scheme (with legend ``Proposed'') to two reactive schemes with and without TTT (with legends ``Reactive (w/~TTT)'' and ``Reactive (w/o~TTT)''). The proactive handover scheme with perfectly predicted next associated SBS is also considered (with legend ``Proactive (PP)''), which provides an upper bound of the handover performance. For a fair comparison, the handover trigger condition is identical for all the proactive and reactive handover schemes.


\begin{figure}[!htb]
	\centering
    \begin{minipage}[t]{0.44\textwidth}
    \includegraphics[width=1\textwidth]{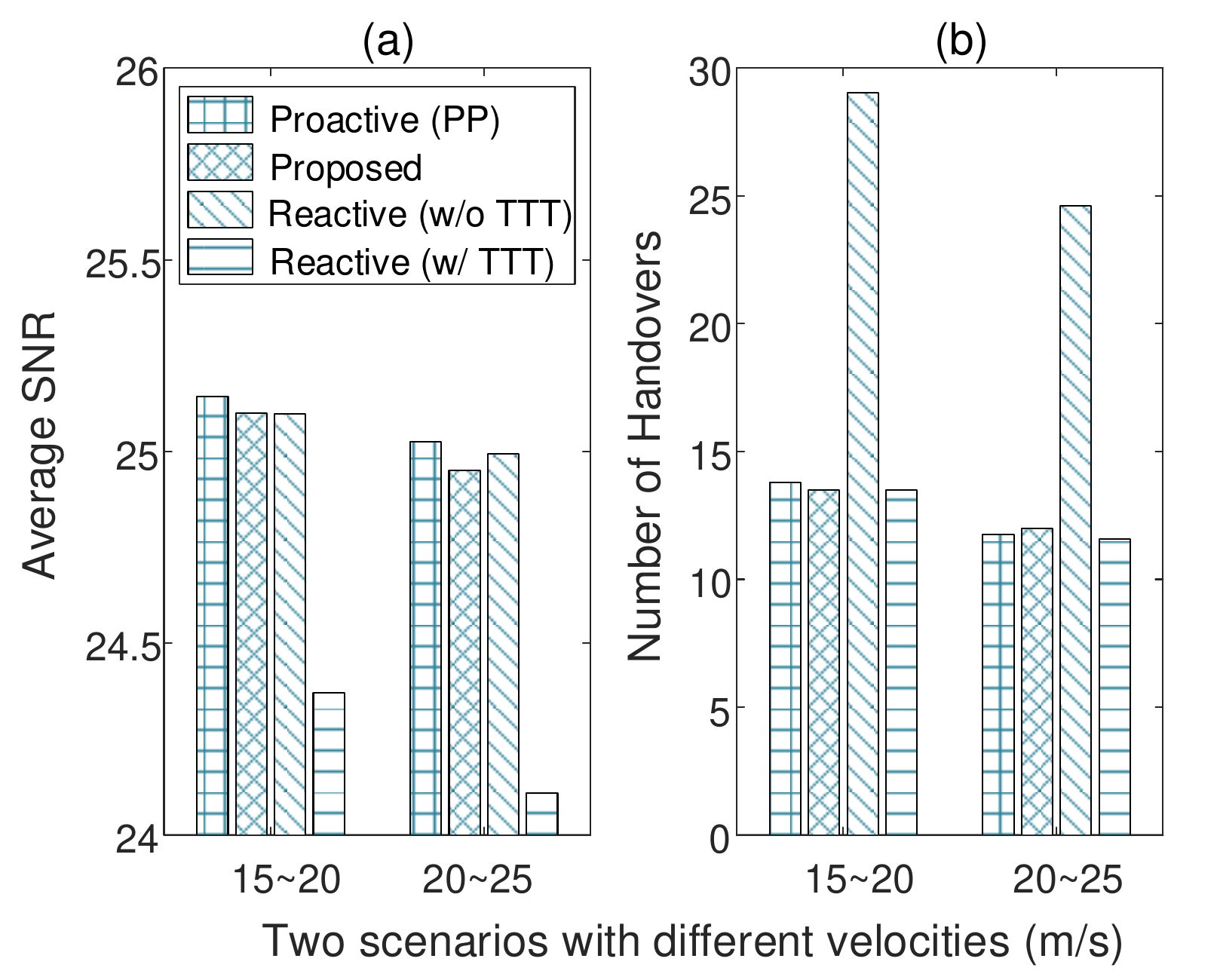}
    \end{minipage}
	\caption{Handover performance comparison of different handover schemes.}
	\label{fig:Handover_performance}
    \vspace{-0.2cm}
\end{figure}

In Fig. \ref{fig:Handover_performance}, we show the handover performance of the proposed handover scheme using the offline-trained model that achieves the target test accuracy of $90\%$ and three baselines. We consider two metrics. One is the \emph{number of handovers}, and the other is the \emph{average SNR} when a user traverses a region, which can reflect the average QoS in this region. Two scenarios with different initial velocities of users are simulated.

We can see that the average SNR and the number of handovers of the proposed handover scheme are very close to those of the ``Proactive (PP)'' scheme.
The proposed handover scheme can increase the average SNR of the user by reducing the handover delay compared to the ``Reactive (w/~TTT)'' scheme, or decrease the number of handovers compared to the ``Reactive (w/o~TTT)'' scheme. This indicates that compared with reactive handover schemes, the proposed handover scheme can reduce the frequent handovers and improve the QoS of users simultaneously.

\section{Conclusions}
In this paper, we investigated the FL for proactive handover in the mmWave vehicular networks. We introduced the streaming FL in the local model update for the users with limited storage space, employed the asynchronous FL in the global model aggregation to increase the number of participant users for improving the learning performance, and considered online training with transfer learning for adapting to the dynamically mobile environment. Simulation results showed that the FL based scheme is with less uplink communication cost than the centralized learning based scheme. Compared to the reactive handover schemes, the proposed handover scheme can effectively reduce the frequent handovers meanwhile improve the QoS of users.

\bibliographystyle{IEEEtran}
\bibliography{ref}
\end{document}